\documentclass[longauth]{aa} 

\usepackage{graphicx}
\usepackage{txfonts}

\usepackage{hyperref}

\newcommand{\hess}{\textsc{H.E.S.S.}}
\newcommand{\fer}{{\sl {\it Fermi}}}
\newcommand{\fla}{\fer-LAT}
\newcommand{\gr}{$\gamma$-ray}
\newcommand{\grs}{$\gamma$-rays}

\usepackage{footmisc}
\makeatletter
\renewcommand*{\fnsymbol}[1]{\ifcase#1\or*\or$\dagger$\or$\ddagger$\or**\or$\dagger\dagger$\or$\ddagger\ddagger$ \fi}
\makeatother

\begin{document}
   \title{The supernova remnant W49B as seen with H.E.S.S. and \fla}

   \authorrunning{The H.E.S.S.~and~Fermi-LAT Collaborations}
   \titlerunning{W49B with \hess\ and \fla}

   \author{\tiny H.E.S.S. Collaboration
     \and H.~Abdalla \inst{1}
     \and A.~Abramowski \inst{2}
     \and F.~Aharonian \inst{3,4,5}
     \and F.~Ait Benkhali \inst{3}
     \and A.G.~Akhperjanian\protect\footnotemark[2] \inst{6,5} 
     \and T.~Andersson \inst{10}
     \and E.O.~Ang\"uner \inst{7}
     \and M.~Arrieta \inst{15}
     \and P.~Aubert \inst{24}
% ...
   }

   \institute{
     Centre for Space Research, North-West University, Potchefstroom 2520, South Africa \and 
     Universit\"at Hamburg, Institut f\"ur Experimentalphysik, Luruper Chaussee 149, D 22761 Hamburg, Germany \and 
     Max-Planck-Institut f\"ur Kernphysik, P.O. Box 103980, D 69029 Heidelberg, Germany \and 
     Dublin Institute for Advanced Studies, 31 Fitzwilliam Place, Dublin 2, Ireland \and 
     National Academy of Sciences of the Republic of Armenia,  Marshall Baghramian Avenue, 24, 0019 Yerevan, Republic of Armenia  \and
     Yerevan Physics Institute, 2 Alikhanian Brothers St., 375036 Yerevan, Armenia \and
     Institut f\"ur Physik, Humboldt-Universit\"at zu Berlin, Newtonstr. 15, D 12489 Berlin, Germany \and
     University of Namibia, Department of Physics, Private Bag 13301, Windhoek, Namibia \and
     GRAPPA, Anton Pannekoek Institute for Astronomy, University of Amsterdam,  Science Park 904, 1098 XH Amsterdam, The Netherlands \and
     Department of Physics and Electrical Engineering, Linnaeus University,  351 95 V\"axj\"o, Sweden \and
     Institut f\"ur Theoretische Physik, Lehrstuhl IV: Weltraum und Astrophysik, Ruhr-Universit\"at Bochum, D 44780 Bochum, Germany \and
     GRAPPA, Anton Pannekoek Institute for Astronomy and Institute of High-Energy Physics, University of Amsterdam,  Science Park 904, 1098 XH Amsterdam, The Netherlands \and
     Institut f\"ur Astro- und Teilchenphysik, Leopold-Franzens-Universit\"at Innsbruck, A-6020 Innsbruck, Austria \and
     School of Physical Sciences, University of Adelaide, Adelaide 5005, Australia \and
     LUTH, Observatoire de Paris, PSL Research University, CNRS, Universit\'e Paris Diderot, 5 Place Jules Janssen, 92190 Meudon, France \and
     Sorbonne Universit\'es, UPMC Universit\'e Paris 06, Universit\'e Paris Diderot, Sorbonne Paris Cit\'e, CNRS, Laboratoire de Physique Nucl\'eaire et de Hautes Energies (LPNHE), 4 place Jussieu, F-75252, Paris Cedex 5, France \and
     Laboratoire Univers et Particules de Montpellier, Universit\'e Montpellier, CNRS/IN2P3,  CC 72, Place Eug\`ene Bataillon, F-34095 Montpellier Cedex 5, France \and
     DSM/Irfu, CEA Saclay, F-91191 Gif-Sur-Yvette Cedex, France \and
     Astronomical Observatory, The University of Warsaw, Al. Ujazdowskie 4, 00-478 Warsaw, Poland \and
     Aix Marseille Universit\'e, CNRS/IN2P3, CPPM UMR 7346,  13288 Marseille, France \and
     Instytut Fizyki J\c{a}drowej PAN, ul. Radzikowskiego 152, 31-342 Krak{\'o}w, Poland \and
     Funded by EU FP7 Marie Curie, grant agreement No. PIEF-GA-2012-332350,  \and
     School of Physics, University of the Witwatersrand, 1 Jan Smuts Avenue, Braamfontein, Johannesburg, 2050 South Africa \and
     Laboratoire d'Annecy-le-Vieux de Physique des Particules, Universit\'{e} Savoie Mont-Blanc, CNRS/IN2P3, F-74941 Annecy-le-Vieux, France
% ...
}

   \offprints{H.E.S.S.~collaboration,
     \protect\\\email{\href{mailto: contact.hess@hess-experiment.eu}{contact.hess@hess-experiment.eu}};
     \protect\\${}^*$ Corresponding authors
     \protect\\${}^\dagger$ Deceased}

  \date{}
   
\abstract{The supernova remnant (SNR) W49B originated from a
  core-collapse supernova that occurred between one and four thousand
  years ago, and subsequently evolved into a mixed-morphology remnant,
  which is interacting with molecular clouds (MC). $\gamma$-ray
  observations of SNR/MC associations are a powerful tool to constrain
  the origin of Galactic cosmic-rays, as they can probe the
  acceleration of hadrons through their interaction with the
  surrounding medium and subsequent emission of non-thermal photons.
  The detection of a $\gamma$-ray source coincident with W49B at very
  high energies (VHE; $E > 100$~GeV) with the H.E.S.S. Cherenkov
  telescopes is reported together with a study of the source with 5
  years of \fla\ high energy $\gamma$-ray (0.06 -- 300~GeV) data. The
  smoothly-connected combined source spectrum, measured from 60~MeV to
  multi-TeV energies, shows two significant spectral breaks at
  $304\pm20$~MeV and $8.4_{-2.5}^{+2.2}$~GeV, the latter being
  constrained by the joint fit from the two instruments. The detected
  spectral features are similar to those observed in several other
  SNR/MC associations and are found to be indicative of $\gamma$-ray
  emission produced through neutral-pion decay.
}


\maketitle

\makeatletter
\renewcommand*{\fnsymbol}[1]{\ifcase#1\@arabic{#1}\fi}
\makeatother


\section{Introduction}\label{intro}

The strong shocks associated with supernova remnants (SNRs) are very
good candidates for accelerating hadronic Galactic cosmic rays to at
least $10^{15}$\,eV through the diffusive shock acceleration mechanism
\citep[e.g.][]{Drury1983}. The detection of very-high energy
\grs\ above $20$\,TeV in objects such as \object{RX J1713.7-3946}
\citep{Aharonian2006b} or \object{RCW 86} \citep{Aharonian2009} shows
that these objects can accelerate particles up to at least
$100$\,TeV. However, whether the bulk of accelerated particles that
radiate in the high energy (HE; $0.1 < E < 100$\,GeV) and
very-high energy (VHE; $E > 100$\,GeV) \gr\ range is of leptonic or
hadronic nature is still under debate \citep{Blasi2013}.

A growing number of SNRs interacting with molecular clouds (SNR/MC)
are being revealed in the GeV and TeV \gr\ domain. This includes
\object{W44} \citep{Abdo2010e}, \object{W28}
\citep{Aharonian2008,Abdo2010c}, \object{CTB 37A}
\citep{Aharonian2008b,Castro2010} and \object{IC 443}
\citep{Albert2007,Acciari2009,Abdo2010d}. Even though isolated SNRs
are obviously cleaner laboratories to study cosmic-ray acceleration
processes, SNR/MC associations are good candidates to test the
presence of hadronic acceleration in SNRs, partly because the HE and
VHE \gr\ emissions from the decay of $\pi^0$ mesons are expected to be
strongly enhanced. The neutral pions, produced when high energy
protons (or nuclei) collide with interstellar material, each decay
into two $\gamma$ rays with equal energies in the pion's
rest-frame. This translates into a steep rise below $\sim$200\,MeV in
the $\gamma$-ray spectral energy distribution (often referred to as
the ``pion-decay bump''). This characteristic spectral feature has
been recently detected at high energies for three interacting SNRs:
IC~443, W44 \citep{Ackermann2013} and \object{W51C}
\citep{Jogler2016}. However, whether this feature is the signature of
the acceleration of freshly injected protons may be questioned as
re-acceleration of diffuse cosmic rays for a limited time period is
also possible \citep{Uchiyama2010, Cardillo2016}.

In this context, the \object{W49} region, discovered in the 22\,cm
survey of \citet{Westerhout1958}, represents one of the most
interesting regions in the Galaxy to study cosmic-ray
acceleration. This region contains two remarkable objects: a young SNR
(\object{W49B}) and a star-forming region (\object{W49A}). The SNR
W49B (G43.3--0.2) is another example of a SNR/MC association. It is a
mixed-morphology \citep{Rho1998} SNR with a bright shell of diameter
$\sim4'$ resolved at radio wavelengths and centrally filled with
thermal X rays \citep{Hwang2000}. With a flux density of $38$\,Jy at
$1$\,GHz, this source is one of the brightest SNRs of the Galaxy at
radio wavelengths. Extensive infrared and X-ray studies revealed that
W49B's progenitor was a supermassive star that created a wind-blown
bubble in a dense molecular cloud in which the explosion occurred
\citep{Keohane2007}. It has recently been shown that W49B's progenitor
experienced a jet-driven core collapse explosion
\citep{Lopez2013,Gonzalez-Casanova2014}. Observations of mid-infrared
lines from shocked molecular hydrogen show that W49B is interacting
with molecular clouds. Near-infrared \ion{Fe}{II} observations
revealed filamentary structures which is evidence of radiative shocks
\citep{Keohane2007}. This interaction is also suggested by \ion{H}{I}
line observations \citep{Brogan2001}. The age of the remnant is
estimated to be between $1000$ and $4000$ years
\citep{Moffett1994,Hwang2000,Zhou2011} and the distance of this object
is still not very well constrained. From \ion{H}{I} absorption
analyses, \citet{Radhakrishnan1972} derived a distance, which
\citet{Moffett1994} later re-calculated to be $\sim 8~\rm kpc$ (using
an updated Galactic rotation model). \citet{Brogan2001}, using more
recent VLA data, have shown that an association of W49B with the
nearby star forming region W49A is also possible, extending the range
of possible distances ($8\leq D \leq 12$\,kpc) for this source. More
recently, \citet{Zhu2014} obtained a distance of $\sim $10\,kpc.

As mentioned above, the other notable component of the W49 region is
W49A \citep{Mezger1967}. It is one of the brightest giant radio
\ion{H}{II} regions in the Galaxy. This star forming region is located
in the densest $\sim$15\,pc of a $10^6\,\rm{M_\odot}$ giant molecular
cloud of $\sim$100\,pc in size. It contains numerous compact and
ultra-compact (UC) \ion{H}{II} regions and its emission is equivalent
to the presence of about $100$ O7 stars \citep{Brogan2001}. The
presence of a very massive star with an initial mass of
$100-180\,\mathrm{M_{\odot}}$ has been reported by
\citet{Wu2014}. W49A is associated with a molecular outflow and strong
$\rm H_2O$ masers \citep{Walker1982,Scoville1986}, the proper motion
of which was used by \citet{Zhang2013} to determine a distance of
$11.11^{+0.79}_{-0.69}$\,kpc. Star forming regions are considered as
potential VHE $\gamma$-ray emitters since they generally host massive
stars which could accelerate particles to VHEs through interactive or
collective wind effects. As an example, the TeV \gr\ source coincident
with Westerlund\,1 \citep{Abramowski2012}, detected with the High
Energy Stereoscopic System (H.E.S.S.), may be the site of such
processes.

In the HE domain, a bright source coincident with W49B is detected
with the \emph{Fermi} Large Area Telescope (LAT) \citep[\object{3FGL
    J1910.9+0906} in the \fla\ 3FGL catalog,][]{3FGL}). It was one of
the 205 most significant sources after the first three months of
observation\footnote{It was designated as \object{0FGL J1911.0+0905}
  in this so-called {\it Bright Sources List} \citep{Abdo2009}.} and
is one of the 360 sources detected above 50~GeV in 80 months of data
\citep{2FHLPaper}. This source is the subject of a detailed
analysis presented in \citet{Abdo2010b}. In that paper, the detection
of HE \gr\ emission was reported towards W49B at a significance of
$38\sigma$ with 17 months of data.  The authors disfavored a possible
pulsar origin for the emission observed towards W49B in the GeV
range. Furthermore, from X-ray measurements, the presence of a neutron
star as the result of the progenitor star's collapse appears unlikely
\citep{Lopez2013}.

In the present paper, the detection of a source coincident with the
SNR W49B in the VHE \gr\ domain with H.E.S.S. is reported. The
analysis of the \fla\ data is applied in this work to 5 years of data
using an updated calibration and updated source and background
models. The spectral and morphological results in the GeV and TeV
regime are discussed in the context of a SNR interacting with a
molecular cloud.


\section{Data analysis}\label{ana}

\subsection{\hess\ observations and analysis}\label{anahess}

\subsubsection{\hess\ data set and analysis methods}
H.E.S.S. is an array of five imaging Cherenkov telescopes located in
the Khomas Highland in Namibia ($23^\circ16'18''$~S,
$16^\circ30'01''$~E), at an altitude of 1800\,m above sea level
\citep{Hinton2004}. The fifth telescope was added in July
2012 but observations with the upgraded system are not used
in this work. With the initial four-telescope array, $\gamma$ rays are
detected above an energy threshold of $\sim$100\,GeV with an angular
resolution better than $0.1^{\circ}$ and an energy resolution below
$15\%$.

A standard data quality selection procedure was used to remove
observations affected by bad weather or instrumental problems
\citep{Aharonian2006}. Extrapolating the \fla\ $\gamma$-ray spectrum
to higher energies, W49B is expected to have a steep spectrum in the
TeV range (this is also a common feature for GeV-detected SNRs
interacting with molecular clouds). Therefore, observations with an
energy threshold above $600\,\mathrm{GeV}$ -- which can be determined
\emph{a priori} from the observation conditions -- were not used in
this study. This resulted in a data set comprising $75$ hours (live
time) of observations towards the W49 region taken from 2004 to 2013
with a mean pointing offset of $1.1^{\circ}$ and a mean zenith angle
of $37^{\circ}$. The data set comprises dedicated observations as well
as observations of other nearby sources or dedicated to the survey of
the Galactic Plane \citep{HGPS2016_2}.

The data were analysed using the Model Analysis \citep{denaurois2009},
in which shower images of all triggered telescopes are compared to a
pre-calculated model by means of a log-likelihood optimisation.  This
advanced reconstruction technique improves the sensitivity with
respect to more conventional ones. The ``Standard cuts'' of the Model
Analysis were adopted. This set of cuts includes a minimum charge in
the shower images of $60$ photoelectrons resulting, for this dataset,
in an energy threshold of $290~\mathrm{GeV}$. This set of cuts was
used for signal extraction as well as for the spectral and
morphological analyses. The results presented in this paper have been
cross-checked with the analysis methods described in \citet{Ohm2009}
and \citet{Parsons2014} and yield compatible results.

\subsubsection{\hess\ analysis results}

\begin{figure}[ht]
  \centering
  \resizebox{\hsize}{!}{\includegraphics{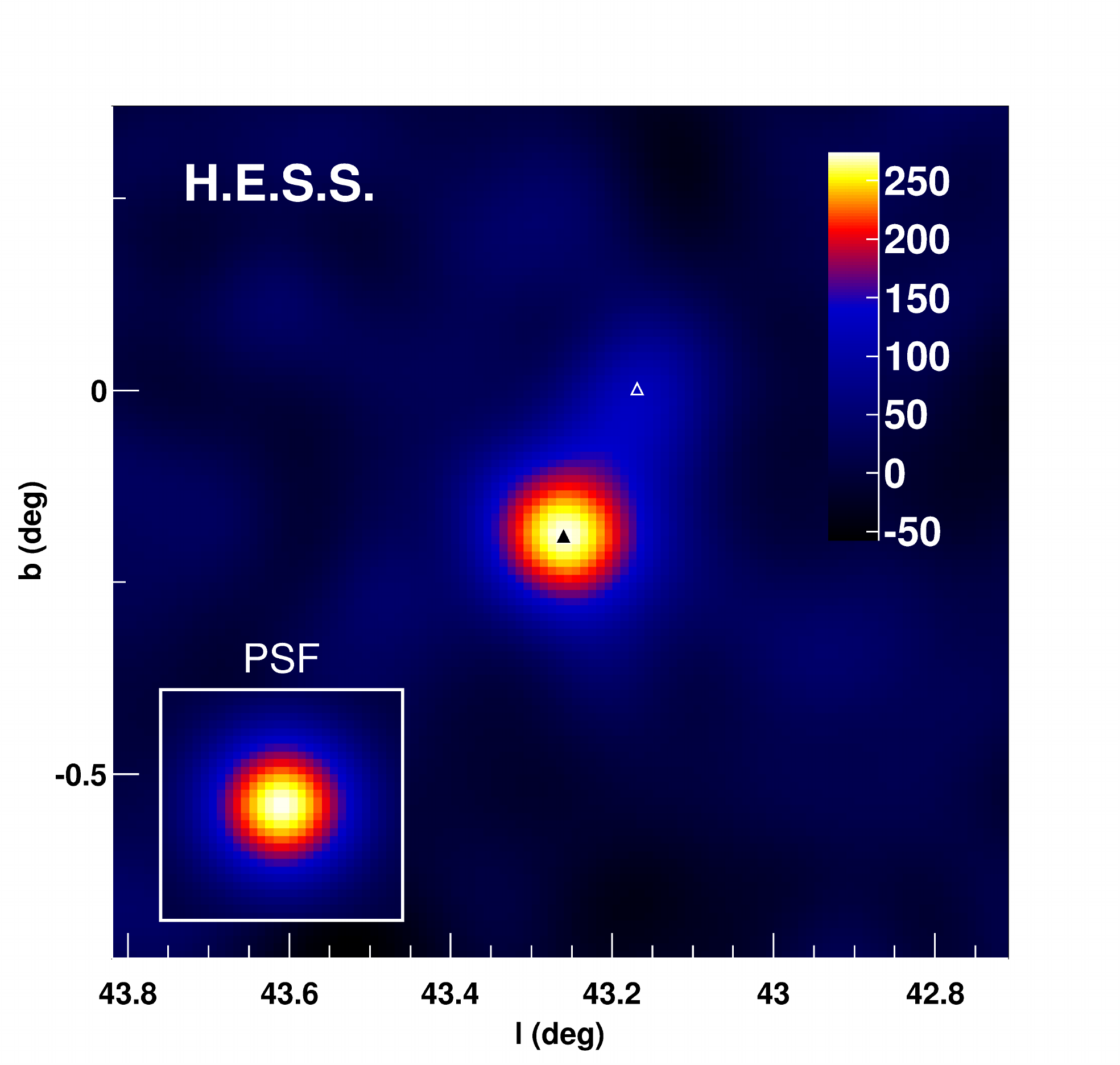}}
  \caption{\gr\ excess map of the W49 region obtained with \hess\ The
    map is smoothed with a Gaussian of width $0.06^{\circ}$,
    corresponding to the 68\% radius containment of the H.E.S.S. point
    spread function (PSF, shown in the inset). The black triangle
    denotes the position of the supernova remnant W49B
    \citep{Green2014}. The white triangle denotes the position of W49A
    \citep{Lockman1989}.}
\label{fig:hess_map1}
\end{figure}

The signal was extracted from a circular region with a radius of
$0.1^{\circ}$ around the position of W49B, taken inside the radio
shell (at the position of the maximum of the emission in X-rays
determined by \citet{Hwang2000}) :
$\left(\ell = 43.275^{\circ}, b = -0.190^{\circ}\right)$.
From the correlated excess map, computed using the ring background
subtraction method \citep{Berge2007} and displayed in
Fig.~\ref{fig:hess_map1}, a significant excess of VHE \grs\ is
detected at the position of W49B, with a peak significance of
$12.9\sigma$ \citep[Eq. 17 of ][]{LiMa1983}. 
VHE $\gamma$-ray emission is observed towards W49A with the primary
analysis but it could not be confirmed (above $5\sigma$) with the
cross-check analysis.  Therefore, only the VHE emission coincident
with W49B is discussed in the following.

\begin{figure}[ht]
  \centering
  \resizebox{\hsize}{!}{\includegraphics{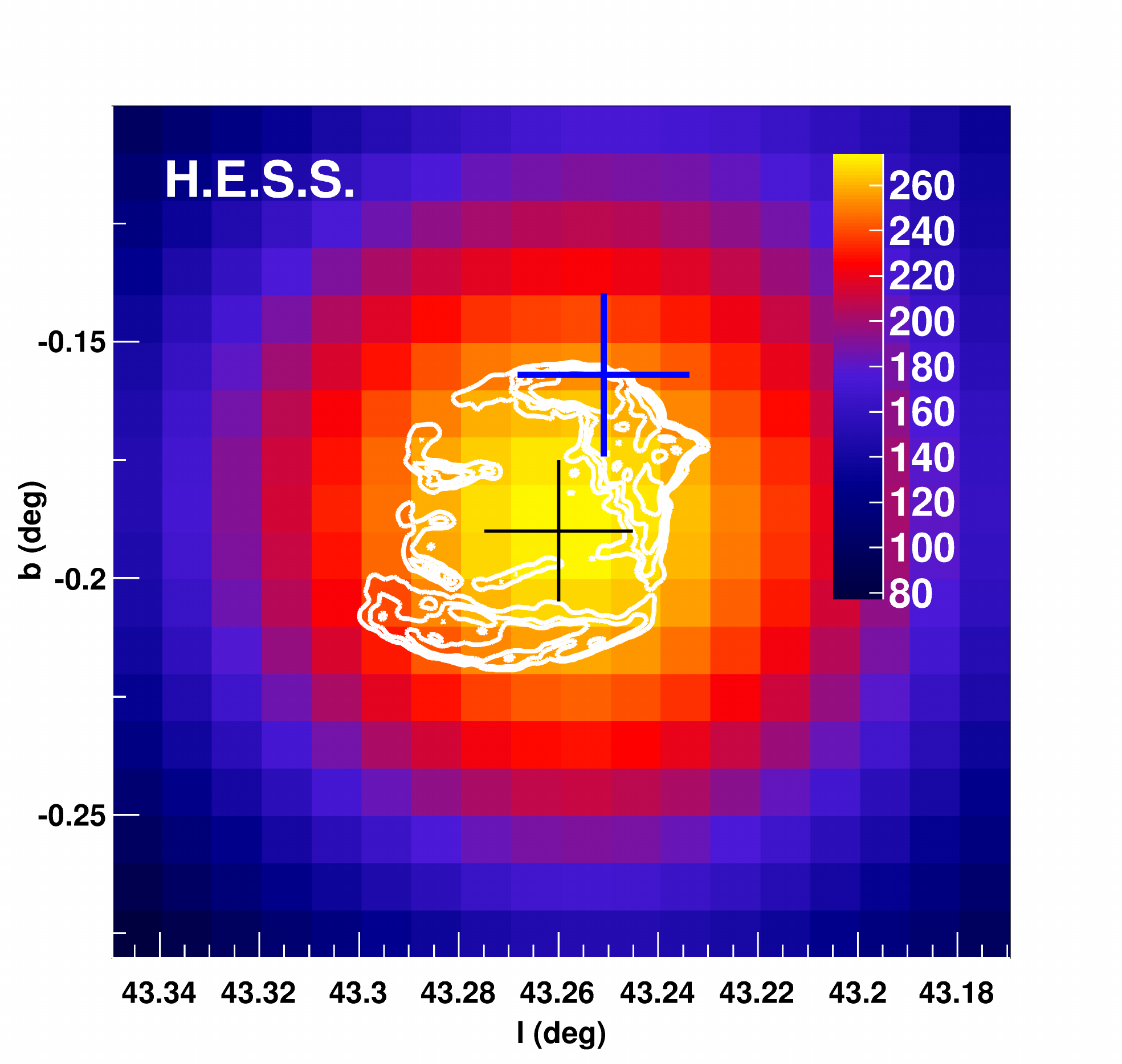}}
  \caption{ Zoomed-in version of Fig.~\ref{fig:hess_map1}. The crosses
    show the best fit positions of the source with \hess\ (black) and
    \fla\ (blue).  The size of the crosses represents the size of the
    95\% Confidence Level (CL) contours on the best-fit positions,
    including systematic uncertainties. The white contours show the
    radio intensity obtained at a wavelength of 20 cm with MAGPIS
    \citep{Helfand2006}.  }
\label{fig:hess_map2}
\end{figure}

In order to derive the source position at VHE and therefore the
best-suited spectral extraction region, the excess map of VHE
\grs\ towards the W49 region was fitted with a two-source
model\footnote{This is done in order to model the region in the best
  possible way, given the possible excess towards W49A which is $\sim
  0.2^{\circ}$ away from W49B (a model with only one source for the
  fit leads to compatible results). The model is convolved with the
  H.E.S.S. point spread function before being fit to the measured
  excess counts.}. The VHE \gr\ emission towards W49B is found to be
point-like with a best-fit position of $\left(\ell =
43.260^{\circ}\pm0.005^{\circ}_{stat}, b = -0.190^{\circ} \pm
0.005^{\circ}_{stat}\right)$ with a systematic error of
$\pm0.006^{\circ}$ on each axis.  As can be seen in
Fig.~\ref{fig:hess_map2}, the obtained position is compatible with the
center of the SNR observed in radio and X-rays
\citep[e.g.][]{Pye1984}. This position corresponds to $\alpha_{\rm
  2000} = 19^{\rm h}11^{\rm m}7.3^{\rm s}$, $\delta_{\rm 2000} =
09^{\circ}05'37.0''$ and the detected source is therefore given the
identifier \object{HESS J1911+090}.

The reflected background subtraction method \citep{Berge2007} was used
at the best fit position to derive the spectrum of the VHE emission
coincident with W49B. From a region of radius $0.1^{\circ}$ around
this position, the total number of counts is 1141, while 16017 events
are used to estimate the background with an average normalisation
factor $\alpha = 0.047$. This corresponds to an excess of $392.3$
events detected with a significance of $13.0\sigma$.  The differential
energy spectrum is derived above $290\; \mathrm{GeV}$ using a forward
folding method described in \citet{Piron2001}. The data are well
described ($\chi^2/\mathrm{ndf} = 60.6/61$) by a power-law model
$dN/dE=\Phi_0 \left(E/1~\mathrm{TeV}\right)^{-\Gamma}$ with a flux
normalisation $\Phi_0 =
\left(3.15\pm0.46_{\rm{stat}}\pm0.63_{\rm{sys}}\right)\times
10^{-13}~\rm{cm}^{-2}~\rm{s}^{-1}~\rm{TeV}^{-1}$ and a spectral index
of $\rm \Gamma = 3.14\pm0.24_{\rm{stat}}\pm0.10_{\rm{sys}}$ (see
Fig.~\ref{fig:fermi_hess_spec}). The integrated flux above 1 TeV is
$F(E > 1\;\mathrm{TeV}) =
\left(1.47\pm0.38_{\rm{stat}}\pm0.29_{\rm{sys}}\right)\times
10^{-13}~\rm{cm}^{-2}~\rm{s}^{-1}$, corresponding to $0.65\%$ of the
Crab Nebula flux above the same energy \citep{Aharonian2006}. Two
other spectral shapes were tested (log parabolic power-law and
power-law with exponential cutoff) but they do not fit the data
significantly better.

\subsection{\fla\ analysis}\label{anafermi}

\subsubsection{$Fermi$-LAT observations and data reduction}

The LAT detects $\gamma$-ray photons by conversion into
electron-positron pairs in the energy range between 20 MeV to higher
than 300 GeV, as described by~\cite{Atwood2009}. It contains a
high-resolution converter/tracker (for direction measurement of the
incident $\gamma$ rays), a CsI(Tl) crystal calorimeter (for energy
measurement), and an anti-coincidence detector to identify the
background of charged particles. The following analysis was performed
using 5 years of \emph{Fermi}-LAT data collected primarily in survey
mode (2008 August 04 -- 2013 August 04). This new dataset consists of
Pass 7 LAT data that have been reprocessed using updated calibration
constants for the detector systems \citep{Bregeon2013}. The primary
differences with respect to the Pass7\_V6 instrument response
functions (IRFs) are the correction of a slight (1\% per year)
expected degradation in the calorimeter light yield and significant
improvement of the position from the calorimeter reconstruction, which
in turn significantly improves the LAT point-spread function above 5
GeV. Only $\gamma$ rays in the Source class events were selected,
excluding those coming from a zenith angle larger than 100$^{\circ}$
to the detector axis to reduce $\gamma$-ray contamination from protons
interacting with the Earth's atmosphere. The $15^{th}$ version of the
reprocessed Pass 7 IRFs for this class of events
\citep[P7REP\_SOURCE\_V15,][]{Bregeon2013} were adopted.

Two different tools were used to perform the spatial and spectral
analysis: $\mathtt{gtlike}$ and
$\mathtt{pointlike}$. $\mathtt{gtlike}$ is a binned maximum-likelihood
method \citep{Mattox1996} implemented in the Science Tools distributed
by the $Fermi$ Science Support Center (FSSC)\footnote{More information
  about the performance of the LAT can be found at the FSSC
  ($\mathtt{http://fermi.gsfc.nasa.gov/ssc}$).}.  $\mathtt{pointlike}$
is an alternate binned likelihood technique, optimized for
characterizing the extension of a source, something that is not
implemented in $\mathtt{gtlike}$. $\mathtt{pointlike}$ was extensively
tested against $\mathtt{gtlike}$ \citep{Lande2012}. These tools fit a
source model to the data along with models for isotropic $\gamma$-ray
emission that also accounts for the residual charged particles and a
diffuse $\gamma$-ray emission model. The obtained log likelihood is
then compared between different source model assumptions to find an
optimal description of the region. For quantitative evaluation between
models, the test-statistic value ($TS$) was defined as $2\log
(\mathcal{L}_1 /\mathcal{L}_0 )$, where $\mathcal{L}_{1/0}$
corresponds to the likelihood value for the source/no-source
hypothesis~\citep{Mattox1996}. The TS value is a measure of the
detection significance or description improvement.

\begin{figure}[ht]
  \centering
  \resizebox{\hsize}{!}{\includegraphics{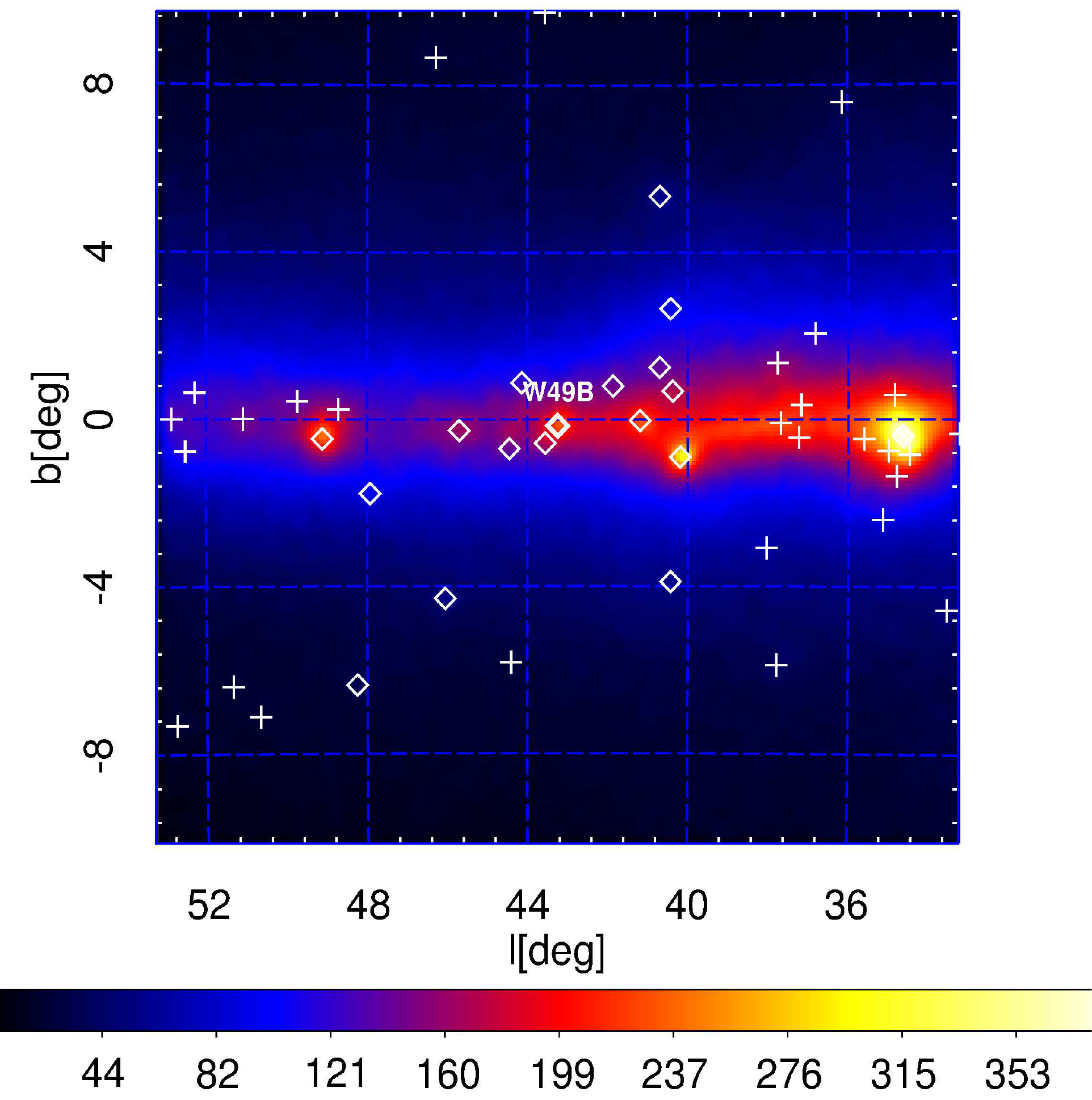}}
  \caption{\emph{Fermi}-LAT count map of the 20$\degr \times$
    20$\degr$ around W49B between 60 MeV and 300 GeV, smoothed with a
    gaussian of width $0.3^{\circ}$. Spectral parameters of sources
    indicated by a diamond are left free in the analysis while sources
    indicated with crosses are frozen in the likelihood fit. Events
    were spatially binned in pixels of side-length $0.1\degr$. The
    dominant source seen at a Galactic longitude of $34.6\degr$ in
    this counts map is W44.}
\label{fig:fermi_counts}
\end{figure}

For this work, the Galactic diffuse emission model
\verb|gll_iem_v05_rev1.fits| and isotropic diffuse model
\verb|iso_source_v05.txt| provided by the \emph{Fermi}-LAT
collaboration were used\footnote{These models are available at:
  http://fermi.gsfc.nasa.gov/ssc/data/access/lat/BackgroundModels.html}. A
20$\degr \times$ 20$\degr$ region of interest binned in 0.1$\degr$
bins and centered on the position of W49B, was defined. Sources within
25$^{\circ}$ of W49B were extracted from the \emph{Fermi}-LAT 3FGL
catalog \citep{3FGL} and included in the likelihood fit. The spectral
parameters of sources closer than 5$^{\circ}$ to W49B or with a TS
larger than 400 were left free and shown with diamonds in
Fig.~\ref{fig:fermi_counts}, while the parameters of all other sources
were fixed at the values from the 3FGL catalog.

\begin{figure}[ht]
  \centering
  \resizebox{\hsize}{!}{\includegraphics{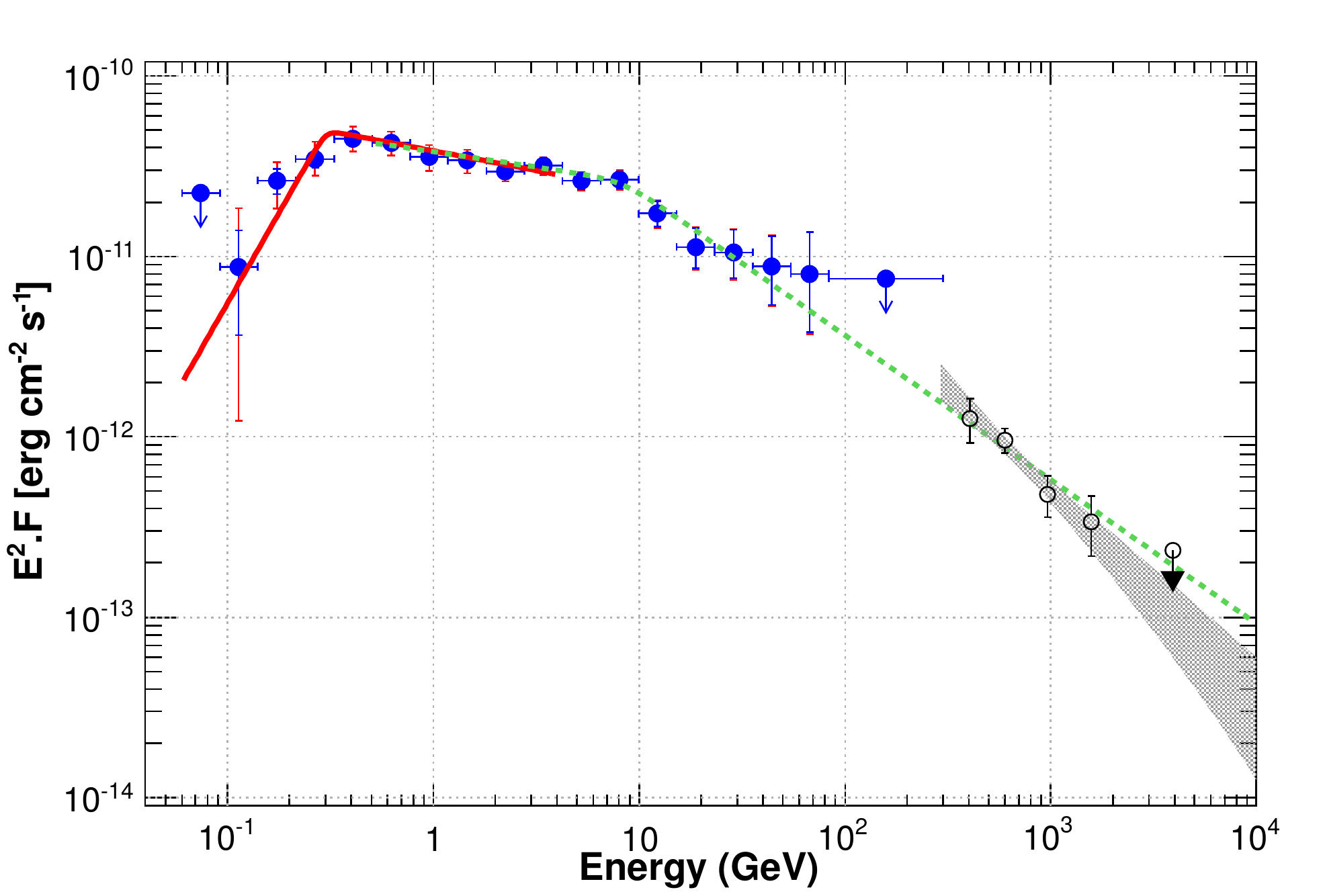}}
  \caption{ \emph{Fermi}-LAT and H.E.S.S. spectrum of W49B. The red
    line shows the best fit of a smoothly broken power-law derived
    between 60 MeV and 4 GeV and the blue data points indicate the
    fluxes measured in each energy bin with the \fla. The statistical
    errors are shown in blue, while the red lines take into account
    both the statistical and systematic errors as discussed in
    Sect.~\ref{fermi:spec}.  The gray band shows the 68\% confidence
    level (CL) uncertainty of the best-fit power-law model with
    H.E.S.S. The open black circles are the spectral points computed
    from the forward-folding fit with their statistical errors shown
    in black.  For both instruments, a 95\% CL upper limit is computed
    when the statistical significance is lower than $2\sigma$.  The
    dotted green line shows the best smoothly broken power-law model
    obtained from the joint fit of the \fla\ and \hess\ data between
    500 MeV and 10 TeV, as described in
    Sect.~\ref{anafermihess_comparison}.}
\label{fig:fermi_hess_spec}
\end{figure}

Two main systematic errors have been taken into account: imperfect
modeling of the Galactic diffuse emission and uncertainties in the
effective area calibration. The first was estimated by comparing the
results obtained using the standard Galactic diffuse model with the
results based on eight alternative interstellar emission models (IEMs)
as performed in \cite{dePalma2013}. We note that the alternative IEMs
do not necessarily bracket the standard IEM but provide a good insight
into deviations from fit results obtained by the standard IEM expected
due to variations in the IEM. Comparing results between analyses
performed using the standard and alternative IEMs is currently the
best method to estimate systematic uncertainties related to the IEM
modeling.  The second major systematic uncertainty was estimated by
using modified IRFs whose effective areas bracket the nominal
ones. These bracketing IRFs are defined by envelopes above and below
the nominal energy dependence of the effective area by linearly
connecting differences of (10\%, 5\%, 5\%, 15\%) at $\log_{10}(E/{\rm
  MeV})$ of (2, 2.5, 4, 6),
respectively\footnote{http://fermi.gsfc.nasa.gov/ssc/data/analysis/LAT\_caveats.html}.

\subsubsection{$Fermi$-LAT analysis results}
\label{fermi:spec}

The spatial analysis was performed using all events above 1~GeV, thus
providing a good compromise between statistics and angular resolution.
The source position, determined by $\mathtt{pointlike}$ is
$\left(\ell = 43.251^{\circ}, b = -0.157^{\circ}\right)$ with a
statistical uncertainty of $0.015^{\circ}$ (95\% CL), compatible with
the radio shell of W49B (see Fig.~\ref{fig:hess_map2}). At this
position, the TS value above 1 GeV is 1504. No significant extension
was found with $\mathtt{pointlike}$.

The spectral analysis was then performed with $\mathtt{gtlike}$
including all events between 0.06 and 300 GeV. Here the best fit
point-source position for W49B obtained by $\mathtt{pointlike}$ was
used in the log likelihood fit. Figure~\ref{fig:fermi_hess_spec} shows
the spectral energy distribution obtained for W49B through the maximum
likelihood estimation. To derive the flux points, the 0.06~--~300 GeV
range was divided into 20 logarithmically-spaced energy bins and a
maximum likelihood spectral analysis was performed in each interval,
assuming a power-law shape with a fixed spectral index of $\Gamma$ = 2
for W49B. The normalizations of the diffuse Galactic and isotropic
emission were left free in each energy bin as well as those of sources
located within $5^{\circ}$ from W49B. A 95\% CL upper limit is
computed when the statistical significance is lower than $2\sigma$. At
the highest energies, bins containing no significant signal were
combined to compute a meaningful upper limit. The errors on the
spectral points represent the statistical and systematic uncertainties
added in quadrature.

The spectrum below 300 MeV is steeply rising, clearly exhibiting a
break at around 300 MeV. To quantify the significance of this spectral
break, a log likelihood fit of W49B between 60 MeV and 4 GeV was
performed -- below the high-energy break previously found in the
1-year spectra \citep{Abdo2010b} -- with both a single power law of
the form $F(E) = \Phi_0 (E/E_0 )^{-\Gamma_1}$ and a smoothly broken
power law of the form $F(E) = \Phi_0 (E/E_0) ^{-\Gamma_1} (1 +
(E/E_{\rm br})^{(\Gamma_2 - \Gamma_1)/\alpha})^{-\alpha}$ where
$\Phi_0$ is the differential flux at $E_0 = 200$ MeV and $\alpha =
0.1$, as was done previously for the cases of IC~443 and W44
\citep{Ackermann2013}. The improvement in log likelihood when
comparing the broken power law to a single power law corresponds to a
formal statistical significance of $8\sigma$ for the low-energy break
(TS of 84 for 2 additional degrees of freedom), as reported in
Table~\ref{table:lowbreak}.  Another fit in the same energy band using
a log-parabola was also performed, showing that the log-parabola
improves the likelihood by only $5\sigma$ (TS of 28 for 1 additional
degree of freedom).  Therefore the best spectral shape for this energy
band is the broken power law.

\newcommand*{\MyIndent}{\hspace*{0.2cm}}%
\begin{table*}[!ht] 
\caption{Spectral parameters in the energy ranges of 60 MeV -- 4 GeV
  and 500 MeV -- 300 GeV for power-law (PL) and smoothly broken
  power-law (SBPL) models, using \fla~data and 290 GeV -- 10 TeV with
  \hess~Systematic errors are provided in parentheses. The reference
  energy for which the $\Phi_0$ is given is 1~TeV for \hess~and
  200~MeV for \fla~and the combined spectrum results.  The last column
  indicates the source TS as described in section
  \ref{anafermi}. }
\centering \resizebox{\hsize}{!}{
\begin{tabular}{lccccc}
\hline
\hline
Model & $\Phi_0$ & $\Gamma_1$ &  $\Gamma_2$ & $E{\rm br}$ & Source TS\\
\hline \\
\fla: & & & & & \\
\MyIndent 60 MeV -- 4 GeV: & $\rm  (10^{-10} cm^{-2}\  s^{-1}\ MeV^{-1})$ & & & (MeV) & \\
\MyIndent PL & $6.2 \pm 0.3$ & $2.07 \pm 0.04$ & $\cdots$ & $\cdots$  &  906 \\
\MyIndent SBPL & $3.3 \pm 0.5 \, (\pm 0.6)$  & $0.10 \pm 0.30 \, (\pm 0.28)$ & $2.21 \pm 0.05 \, (\pm 0.06)$ & $304 \pm 20 \, (\pm 10)$  &  990\\
&&&&& \\
\cline{3-4} \\
\MyIndent 500 MeV -- 300 GeV: & $\rm  (10^{-10} cm^{-2}\  s^{-1}\ MeV^{-1})$ & & & (GeV) & \\
\MyIndent PL & $10.4 \pm 0.8$ & $2.32 \pm 0.03$ & $\cdots$ & $\cdots$  &  1570 \\
\MyIndent SBPL & $8.2 \pm 1.2 \, (\pm 2.3)$ & $2.20 \pm 0.06 \, (\pm 0.16)$ & $2.89\pm 0.23 \, (\pm 0.15)$ & $9.4 \pm 2.9 \, (\pm 3.8)$  &  1584 \\ 
&&&&& \\
\hline
\hline \\
\hess: & & & & & \\
\MyIndent >290 GeV: & $\rm  (10^{-13} cm^{-2}\  s^{-1}\ TeV^{-1})$ & & & & \\
\MyIndent PL & $3.15 \pm 0.46 \, (\pm 0.63)$ & $3.14 \pm 0.24 \, (\pm 0.10)$ & $\cdots$ & $\cdots$ & \\
&&&&&\\
\hline 
\hline \\
\fla~\&~\hess : & & & & & \\
\MyIndent 500 MeV -- 10 TeV: & $\rm  ( 10^{-10} cm^{-2}\  s^{-1}\ MeV^{-1})$ & & & (GeV)& \\
\MyIndent SBPL & $7.5 \pm 1.2 $ & $2.17 \pm 0.06$ & $2.80 \pm 0.04$ & $8.4_{-2.5}^{+2.2}$ \\
\\
\hline
\hline
\end{tabular}
}
\label{table:lowbreak}
\end{table*}

\subsection{\hess\ and \fla\ combined results}\label{anafermihess_comparison}

The best-fit position obtained with the Fermi-LAT is coincident with
the radio shell while the fitted position in the TeV range is more
compatible with its center (see Fig.~\ref{fig:hess_map2}). However,
the size of the shell detected at radio wavelengths is comparable to
the size of the H.E.S.S. point spread function, preventing the
discrimination between the emission originating from the center of the
remnant or from its shell. Furthermore, the spectra derived in the GeV
and TeV energy ranges are in good agreement (see
Fig.~\ref{fig:fermi_hess_spec}). The positions and spectra are also
compatible with those reported in the recently published second
\fla\ catalog of sources above 50~GeV \citep{2FHLPaper}. It is
therefore reasonable to assume that the detected $\gamma$-ray photons
are emitted in the same physical region of the W49B remnant.

Taking advantage of the smooth connection between the HE and VHE
spectra, the presence of another break at high energy was evaluated
through a likelihood ratio test statistic applied to both the
H.E.S.S. and \emph{Fermi}-LAT data. For this test, a single power-law
and a smoothly broken power-law between 500 MeV and 10 TeV -- above
the low energy break found above -- were used and the statistical and
systematic uncertainties have been taken into account. The minimum of
the likelihood ratio is reached at 8.4 GeV with a $6.5\sigma$
improvement of the smoothly broken power-law model with respect to the
power-law one. The 68\% confidence interval is [5.9; 10.6] GeV and the
spectral indices below and above the break are respectively $2.17 \pm
0.06$ and $2.80 \pm 0.04$. This result has been checked by fitting the
\fla\ fluxes of W49B between 500 MeV and 300 GeV with both a single
power-law and a smoothly broken power-law. A $3.3\sigma$ (TS of 14 for
2 additional degrees of freedom) improvement is obtained for an energy
break at $9.4 \pm 2.9_{stat} \pm 3.8_{sys}$ GeV. This energy, as well
as the spectral indices above and below the break (listed in Table
\ref{table:lowbreak}), are compatible with the values obtained from
the \fla\ and \hess\ joint fit.


\section{Discussion of the results}\label{discussion}

\subsection{General considerations}
\label{sec:Interp}

A spectral break is detected at $304\pm20$\,MeV with the Fermi-LAT,
with properties (break energy, spectral indices) similar to those
detected in the SNRs IC~443 and W44 \citep{Ackermann2013} and
interpreted as the signature of $\gamma$-ray emission produced through
neutral-pion decay. In addition, the \hess\ and \fla\ joint spectral
fit allowed for the estimation of a second spectral break at
$8.4_{-2.5}^{+2.2}$ GeV, compatible with the break energy position
found using the \emph{Fermi}-LAT data alone ($9.4 \pm 2.9$ GeV). Such
a spectral break has been found in several other interacting SNRs, as
for instance in W51C \citep{FermiW51C}, W28 \citep{Abdo2010c} or
\object{G349.7+0.2} \citep{HessG349}. The energy at which this break
occurs is found to be in the range $1-20$ GeV, except for G349.7+0.2
where it is found at a higher energy ($\sim 50$ GeV). Interestingly,
while W49B is estimated to be $1-4$ thousand years old, it exhibits
spectral features similar to the other detected SNR/MC systems which
are believed to be $\sim 10$ times older. Likewise, G349.7+0.2 which
has roughly the same age as W49B, shows an energy break which is $\sim
5$ times higher. This emphasises the importance of the environment in
the evolution of the $\gamma$-ray luminosity of these objects.

The origin of the break in the $1-20$ GeV range has been the subject
of several theoretical works. It could be explained either by the
diffusion of particles escaping the shock
\citep[e.g.][]{Ohira2011,Li2012}, by acceleration and wave evanescence
effects \citep[e.g.][]{Inoue2010,Malkov2012}, or time-limited
re-acceleration of CRs already present in the surroundings
\citep[e.g.][]{Uchiyama2010,Cardillo2016}. Other phenomena might be at
play in such objects, such as a so-called neutral return flux that can
modify the dynamics of the shock when propagating into a partially
neutral medium \citep{Blasi2012}. The maximum energy that accelerated
particles can achieve could be limited by the ion-neutral damping
\citep{PtuskinZirakashvili2005} and their spectrum could be steepened
due to the feedback of amplified magnetic fields
\citep{Ptuskin2010,Caprioli2012}. Future precision modeling may be
able to take the whole history and environment of W49B into account to
predict the spectra of accelerated particles in this source and
resultant radiation from radio to TeV energies. In this work, a simple
one-zone model that assumes an empirical distribution of accelerated
particles is used specifically to constrain the energetics and the
dominant processes in the source.

\subsection{Molecular gas density estimates}
\label{sec: density}

In order to use appropriate values for the modeling, the mass and
density of the molecular gas inside a spherical volume of radius
$0.025^{\circ}$, consistent with that of the radio SNR radius, were
estimated using the Galactic Ring Survey $^{13}\rm{CO}$ data
\citep{Jackson2006}. For the three velocity ranges (1-15, 38-47 and
57-67\,km/s at their kinematic distances of 11, 9.3 and 5\,kpc
respectively) examined by \citet{Zhu2014} (see also \citet{Chen2014}),
and using the $^{13}\rm{CO}$ ``X-factor'' of
$4.92\times10^{20}\,\rm{(K\,km/s)}^{-1}$ from \citet{Simon2001},
masses of $5\times10^{3}$ to $17\times10^{3}\,\rm{M_{\odot}}$ with
densities of 400 to 4\,000\,cm$^{-3}$ were obtained. Allowing for a
larger radius to the outer edge of the radio SNR features
($0.05^{\circ}$) reduces the molecular gas densities by a factor
$\sim$\,2 to 3.

One can note that \citet{Chen2014} and \citet{Zhu2014} have argued
that the CO bubble in the 38-47\,km/s range was blown out by the W49B
progenitor star, suggesting a distance of about 10 kpc for the
SNR. Additionally for this, and the 57-67 km/s ranges, the $\rm{H}_2$
column densities (about $5\times10^{22}\,\rm{cm}^{-2}$) inferred in
this work are consistent with that derived from the X-ray spectral
studies of \citet{Keohane2007}.

\subsection{High-Energy Particles in W49B}
\label{sec: HEP}

The observed $\gamma$-ray emission is most likely to originate from
SNR W49B, as discussed in \cite{Abdo2010b}.  In this section, the
properties of the relativistic particles that radiate at $\gamma$-ray
energies are investigated.  Given the smooth connection of the
observed \fla\ and H.E.S.S. spectra (see
Fig.~\ref{fig:fermi_hess_spec}), it is assumed that the $\gamma$-ray
photons are emitted in the same region.  In addition, the observed
radio emission\,\citep[e.g.][]{Moffett1994} is assumed to come from
the same shell region where a constant hydrogen density $n_{\rm H}$
and magnetic field strength $B$ are considered.  Given the evidences
that W49B interacts with the surrounding molecular
clouds\,\citep[e.g.,][]{Keohane2007,Zhu2014}, we consider a scenario
where the $\gamma$-ray emission comes from the shocked molecular
clouds in the SNR shell.  Note that the emitting region may originate
from swept-up stellar wind and/or interstellar gas as discussed in
\cite{Abdo2010b}. However, adopting such a scenario would not
significantly change the spectra calculated in this section.

The $\gamma$-ray emission is reproduced using the (relatively) simple
model described in \citet{FermiW51C}.  Empirical injection
distributions of protons and electrons are assumed:

\begin{equation}
Q_{e,p}(p)=a_{e,p}\left(\frac{p}{p_0}\right)^{-s}\left(1+\left(\frac{p}{p_{\rm br}}\right)^2\right)^{-\Delta s/2},  
\end{equation}
where $p_0 = 1\ {\rm GeV}\, c^{-1}$, $p_{\rm br}$ is the break
momentum, $s$ is the spectral index below $p_{\rm br}$ and $\Delta s$
is the difference between the spectral indices values below and above
$p_{\rm br}$. The kinetic equation for the momentum distribution of
high-energy particles in the shell can be written as:

\begin{equation}
\frac{\partial N_{e,p}}{\partial t}  = \frac{\partial}{\partial p} ( b_{e,p} N_{e,p})  + Q_{e,p},
\end{equation}
where $b_{e,p}=-dp/dt$ is the momentum loss rate (only the radiative
losses -- synchrotron, inverse-Compton, bremsstrahlung and neutral
pion decay -- were considered), and $Q_{e,p}(p)$ (assumed to be
time-independent) is the particle injection rate.  To obtain the
radiation spectrum from the remnant, $N_{e,p}(p, T_0)$ is numerically
calculated for $T_0= 2000$\,yr.

\begin{figure*}[h] 
  \begin{center}
    \resizebox{\hsize}{!}{\includegraphics{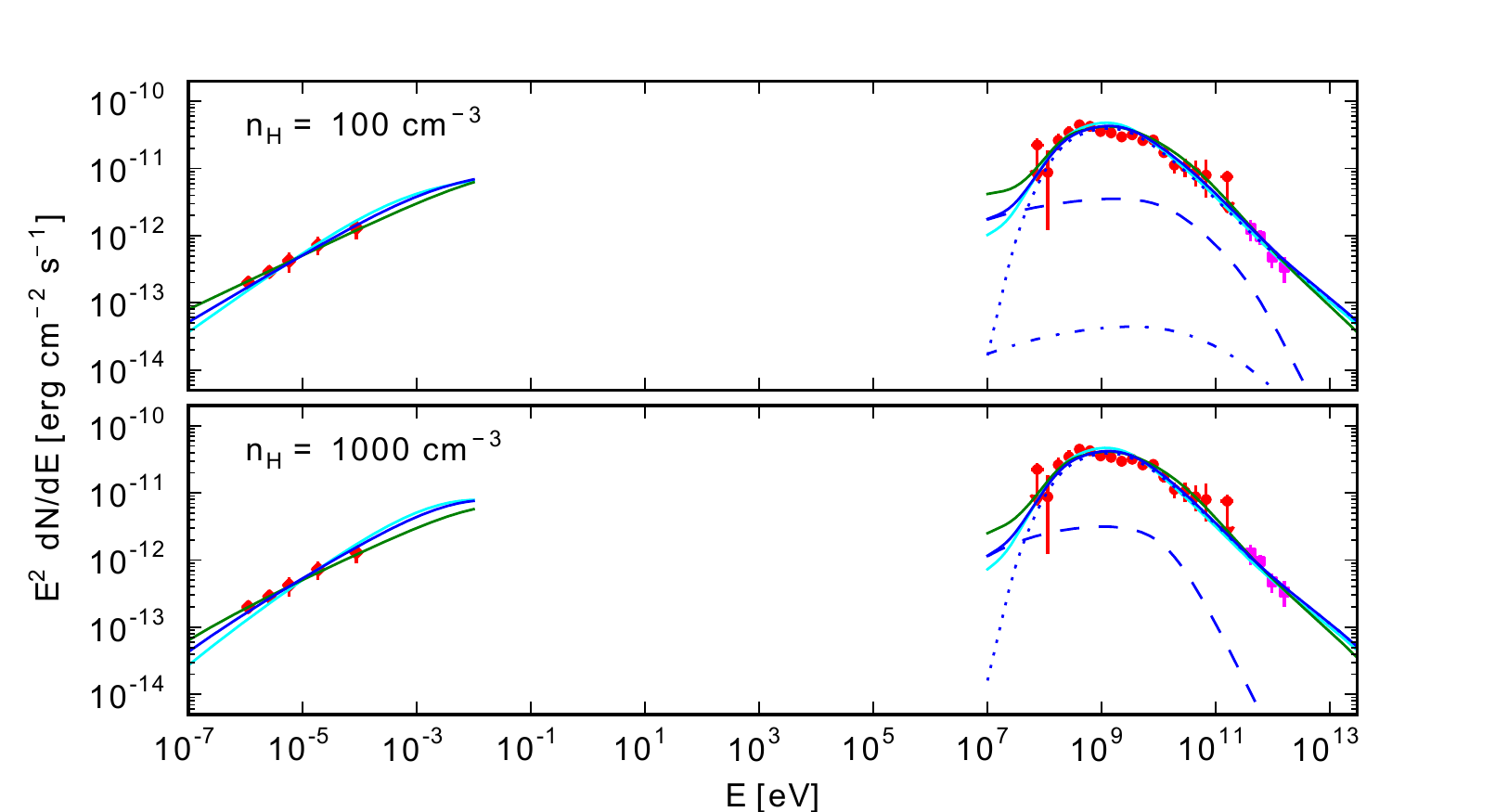}}
    \caption{\small SEDs of W49B with model curves for the
      hadronic-dominant scenario. The upper and the lower panels show
      (a1--3) and (a4--6) respectively (see Table\,\ref{tbl:
        model_had}).  The red diamonds, red circles, and magenta
      squares represent observed data in the
      radio\,\citep{Moffett1994}, LAT, and H.E.S.S. bands
      respectively.  The radio emission is explained by the
      synchrotron radiation from the relativistic electrons.  The
      $\gamma$-ray emission can be decomposed into $\pi^0$-decay
      (dotted line), bremsstrahlung (dashed line), and IC scattering
      (dot-dashed line).  The solid line represents the total flux of
      the components.  The cases (a1)/(a4), (a2)/(a5), and (a3)/(a6)
      are represented by cyan, blue, and green lines respectively in
      the upper/lower panel.  The decomposed emissions are shown for
      the cases (a2) and (a5) in the upper and the lower panels
      respectively.
      \label{fig: model_had}
    }
  \end{center}
\end{figure*}

\begin{figure*}[h] 
  \begin{center}
    \resizebox{\hsize}{!}{\includegraphics{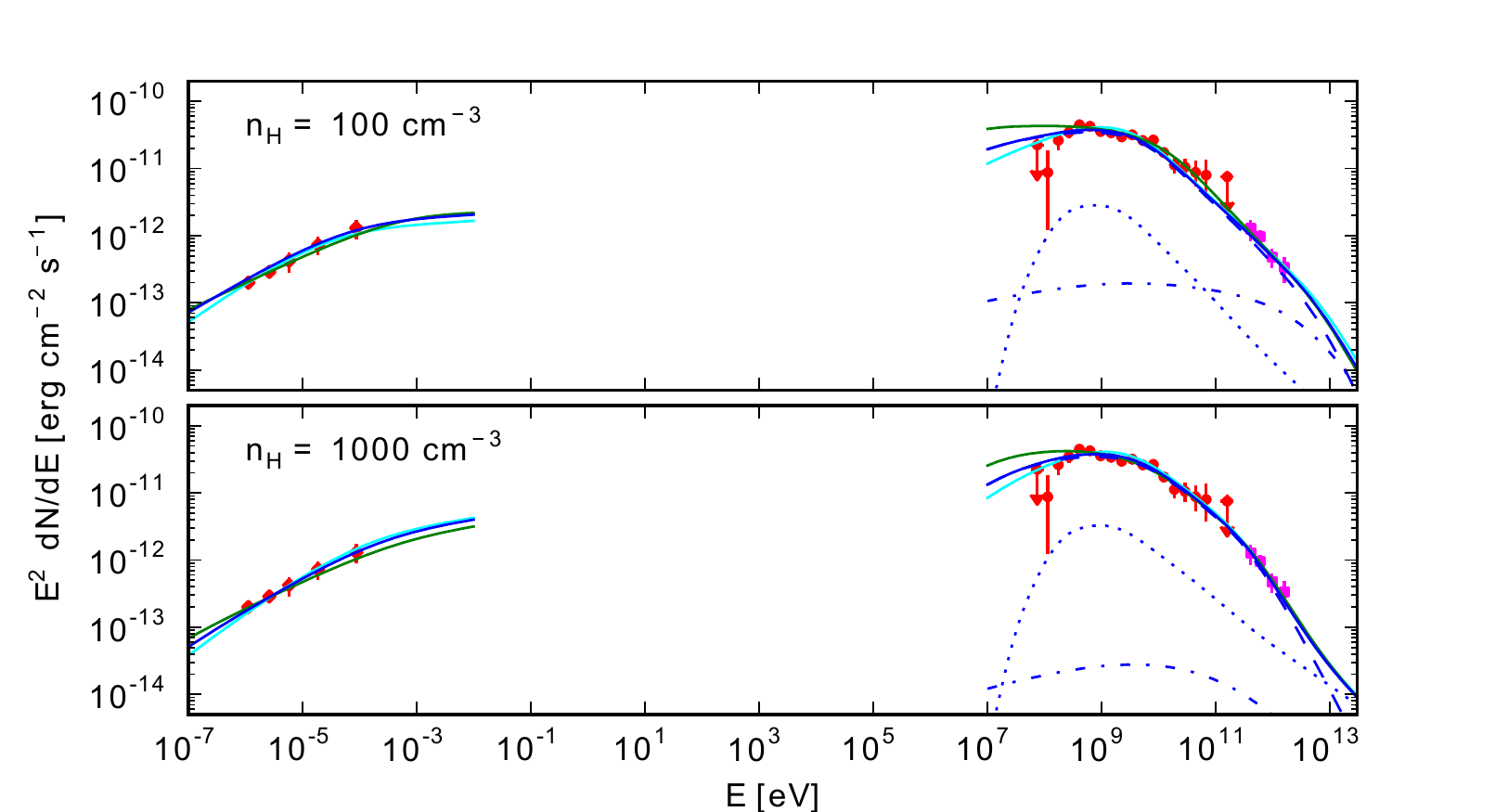}}
    \caption{\small Same as Fig.~\ref{fig: model_had} but for the
      leptonic-dominant scenario.  The upper and the lower panels show
      reproduced spectra of (b1--3) and (b4--6) respectively (see
      Table\,\ref{tbl: model_lep}).  The cases (b1)/(b4), (b2)/(b5),
      and (b3)/(b6) are represented by cyan, blue, and green lines
      respectively in the upper/lower panel.  The decomposed emission
      is shown for the cases (b2) and (b5) in the upper and the lower
      panels respectively.
      \label{fig: model_lep}
    }
  \end{center}
\end{figure*}

The $\gamma$-ray emission mechanisms include the $\pi^0$-decay
$\gamma$ rays due to high-energy protons, bremsstrahlung and
inverse-Compton scattering processes by high-energy electrons.  The
radio emission is calculated by considering the synchrotron radiation
from the population of high-energy electrons.  To reproduce the
multi-wavelength spectra, two cases for the relativistic electrons to
protons number ratio ($K_{ep} \equiv a_e/a_p$) were considered,
$K_{ep} = 0.01$ and $K_{ep} = 1$.  The value of 0.01 is similar to
what is locally observed for cosmic rays at GeV
energies\,\citep[e.g.,][]{Beischer2009} whereas $K_{ep} = 1$
corresponds to an upper limit of the ratio that is most likely too
high for realistic applications due to the faster cooling of the
electrons compared to protons.  In the calculation, the same indices
and break momenta for both the electrons and the protons were chosen.
Given that the radio synchrotron index $\alpha = 0.48$
\citep{Moffett1994} corresponds to $s \sim 2$, three values for $s$
were considered: 1.8, 2.0, and 2.2.  The observed $\gamma$-ray spectra
are calculated by adjusting spectral parameters of the relativistic
particles: the normalization, the break momentum $p_{\rm br}$, and
$\Delta s$.  The magnetic field $B$ is then determined to explain the
observed radio data. Only thermal X-rays, which are not useful in
constraining the non-thermal electron population energy distribution,
are observed from the SNR \citep{Hwang2000}. In the energy range 1--10
keV, our model synchrotron X-ray predictions would be at least one
order of magnitude below the measured thermal X-ray values; thus no
constraining limits exist in this part of the spectrum.  Two
conservative cases of 100\,cm$^{-3}$ and 1000\,cm$^{-3}$ for the
ambient medium density $n_{\rm H}$ were considered (see Sec. \ref{sec:
  density}).  The seed photons for inverse-Compton scattering
processes include infrared ($kT_{\rm IR} = 3\times 10^{-3}$ eV,
$U_{\rm IR} = 0.7\ \rm eV\ cm^{-3}$), optical ($kT_{\rm opt} = 0.3$
eV, $U_{\rm opt} = 0.8\ \rm eV\ cm^{-3}$), and the CMB. These values
were taken to match the axisymmetric estimates of the GALPROP code at
the assumed distance of W49B \citep[][and references
  therein]{Galprop2008}. For a more precise modeling, dedicated
analysis of the infrared and optical emission around the source should
be performed.  The derived parameters and spectra are shown in
Tables\,\ref{tbl: model_had} and \ref{tbl: model_lep}, and
Figs.~\ref{fig: model_had} and \ref{fig: model_lep}.

\begin{table}[ht!]
  \caption{Parameters for the hadronic scenario \label{tbl: model_had}}
  \centering
  \begin{tabular}{lcccccc}
    \hline \hline
    Model & $n_{\rm H}$ & $s$ & $\Delta s$ 
    & $p_{\rm br}$  & $B$ & $W_p$ \\
    & (${\rm cm}^{-3}$) &  &  & (GeV $c^{-1}$)  & ($\mu$G) & ($10^{49}$ erg) \\
    \hline
    (a1) & 100 & 1.8 & 1.0 & 15 & 100 & 12\\
    (a2) & 100 & 2.0 & 0.8 & 30 & 100 & 11\\
    (a3) & 100 & 2.2 & 0.7 & 100 & 90 & 12\\
    (a4) & 1000 & 1.8 & 1.0 & 15   & 500 & 1.1\\ 
    (a5) & 1000 & 2.0 & 0.8 & 30   & 500 & 1.1\\ 
    (a6) & 1000 & 2.2 & 0.7 & 100 & 400 & 1.1\\ 
    \hline
  \end{tabular}
  \tablefoot{ $K_{ep}$ is set to 0.01. The total
    kinetic energies ($E_{\rm kin}$) of the radiating particles
    ($W_{e,p}$) are calculated for $E_{\rm kin} > 100\,{\rm MeV}$ at
    an assumed distance of 10\,kpc.  }
\end{table}

In the case of $K_{ep} = 0.01$ (the hadronic scenario), the obtained
$\gamma$-ray spectrum is dominated by the $\pi^0$-decay emission due
to the relativistic protons as shown in Fig.~\ref{fig: model_had}.
The observed spectra can be formally explained in all the cases except
for the case (a4) where the reproduced synchrotron spectrum seems a
little harder than the observed radio data. The rather high values
derived for the magnetic field of models (a4--6) are needed to explain
the radio data. This is due to the fact that $K_{ep}$ is fixed to
$0.01$, allowing for higher values of $K_{ep}$ would imply higher
total kinetic energies for the electrons and lower magnetic field
values would be needed to explain the radio data.

Note that the reproduced spectral shape is independent of the value of
$n_{\rm H}$, as long as the range is set to reasonable values ($n_{\rm
  H} \sim 100\mbox{--}1000\,{\rm cm}^{-3}$).

\begin{table}[ht!]
  \caption{Parameters for the leptonic scenario \label{tbl: model_lep}}
  \centering 
  \begin{tabular}{lcccccc}
    \hline \hline
    Model & $n_{\rm H}$ & $s$ & $\Delta s$ 
    & $p_{\rm br}$  & $B$ & $W_e$ \\
    & (${\rm cm}^{-3}$) & & & (GeV $c^{-1}$) & ($\mu$G) & ($10^{49}$ erg) \\ 
    \hline
    (b1) & 100      & 1.8 & 1.1 & 7 & 22 & 1.6\\
    (b2) & 100      & 2.0 & 0.9 & 10 & 25 & 1.6\\
    (b3) & 100      & 2.2 & 0.8 & 30 & 22 & 2.0\\
    (b4) & 1000    & 1.8 & 0.9 & 7 & 100 & 0.16\\ 
    (b5) & 1000   & 2.0 & 0.7 & 10 & 100 & 0.16\\
    (b6) & 1000   & 2.2 & 0.5 & 15 & 90 & 0.19\\
    \hline
  \end{tabular}
  \tablefoot{ Same as Table\,\ref{tbl: model_had} except for $K_{ep}$,
    which is set to 1.  }
\end{table}

In the case of $K_{ep} = 1$ (the leptonic scenario), the obtained
$\gamma$ rays are dominated by the bremsstrahlung emission due to the
relativistic electrons as shown in Fig.~\ref{fig: model_lep}.  The
figure shows that $s \lesssim 1.8$ is required to explain the LAT data
below $\sim 300$\,MeV.  The synchrotron radiation loss modifies the
reproduced spectra at the $\gtrsim {\rm TeV}$ range; its break
momentum of the electrons are 16\,TeV\,$c^{-1}$ and 0.6\,TeV\,$c^{-1}$
for $B$ of 20\,$\mu$G (corresponding to the cases (b1--3)) and
100\,$\mu$G (corresponding to the cases (b4--6)) respectively.  On the
other hand, the spectral shape does not significantly change in the
energy band below the break.

As discussed in Sect.~\ref{anafermi}, the LAT data show a clear
spectral break at $304 \pm 20$\,MeV.  Given the systematic errors of
the data, both the hadronic and the leptonic scenarios can explain
this spectral shape.  In the leptonic scenario, however, a limit to
the spectral index of $s \lesssim 1.8$ must be set to reproduce such a
sharp break.  On the other hand, the hadronic scenario can naturally
explain the break by the $\pi^0$-decay bump irrespective of the
spectral index.  The obtained B-field values in Table \ref{tbl:
  model_had} and \ref{tbl: model_lep} are significantly larger than
typical interstellar values. To some extent, these values correspond
to those expected in a MC \citep{Crutcher2010}. They are also
consistent with predictions of magnetic field amplification from
cosmic-ray-excited turbulence \citep{Bell2004,Zirakashvili2008}. The
values adopted in this modeling are set to reasonable values and
different assumptions for the parameters (age, distance, $n_{\rm H}$,
$K_{ep}$) would slightly change the quantitative results of the
modeling (derived magnetic field or energy values) but would not
change these conclusions.

\section{Conclusions}\label{conclusion}

A source of very-high energy $\gamma$ rays is detected towards the
supernova remnant W49B with H.E.S.S. and a joint study is conducted
together with the analysis of 5 years of Fermi-LAT data towards the
source. The point spread functions of these instruments are comparable
to the physical size of the SNR and do not allow for detailed
morphology studies. However, the significantly increased data set
compared to previous publications allowed for the derivation of the
$\gamma$-ray spectrum of the source between 60 MeV and a few TeV. Two
spectral breaks could be identified in the $\gamma$-ray spectrum: one
at 304 MeV and the other at 8.4 GeV. The latter, constrained from a
joint fit of the \fla\ and H.E.S.S. data, is similar to the spectral
breaks observed in other supernova remnants interacting with molecular
clouds and further supports the evidence of interaction observed in
other wavelengths for this object. The broad band spectrum of W49B
could be explained either by leptonic or hadronic models. However, the
sharp break detected with the \fla\ data at 304 MeV seems to favor a
hadronic origin of the $\gamma$-ray emission since the leptonic models
would require an even harder spectrum for the electron population than
the ones tested in this work which would fail to explain the radio
observations.

In the near future, W49B as well as the nearby star forming region
W49A will be of particular interest to study at VHE with improved
analysis methods and with the next generation of instruments such as
the Cherenkov Telescope Array. Observations or analyses of the W49
region with improved angular resolution and higher sensitivity would
help to constrain the morphology and the origin of the emission
towards W49B and possibly confirm the hint of emission towards W49A.
Furthermore, if one assumes that the distances to W49A and W49B are
comparable, then the observed difference between the $\gamma$-ray
luminosities of the two sources would become especially
interesting. It would imply that in the absence of recognizable
supernova remnants -- as in W49A -- the other possible energetic
particle sources like the shocks expected from interacting or
collective stellar winds appear not very effective for HE and VHE
$\gamma$-ray emission in this case. Therefore, a truly reliable
distance determination for these sources (see Sect.~\ref{intro}) would
be of great astrophysical importance.

\begin{acknowledgements}
The support of the Namibian authorities and of the University of
Namibia in facilitating the construction and operation of H.E.S.S. is
gratefully acknowledged, as is the support by the German Ministry for
Education and Research (BMBF), the Max Planck Society, the German
Research Foundation (DFG), the French Ministry for Research, the
CNRS-IN2P3 and the Astroparticle Interdisciplinary Programme of the
CNRS, the U.K. Science and Technology Facilities Council (STFC), the
IPNP of the Charles University, the Czech Science Foundation, the
Polish Ministry of Science and Higher Education, the South African
Department of Science and Technology and National Research Foundation,
the University of Namibia, the Innsbruck University, the Austrian
Science Fund (FWF), and the Autrian Federal Ministry for Science,
Research and Economy, and by the University of Adelaide and the
Australian Research Council. We appreciate the excellent work of the
technical support staff in Berlin, Durham, Hamburg, Heidelberg,
Palaiseau, Paris, Saclay, and in Namibia in the construction and
operation of the equipment. This work benefited from services provided
by the H.E.S.S. Virtual Organisation, supported by the national
resource providers of the EGI Federation.\\

\indent The \textit{Fermi} LAT Collaboration acknowledges generous
ongoing support from a number of agencies and institutes that have
supported both the development and the operation of the LAT as well as
scientific data analysis.  These include the National Aeronautics and
Space Administration and the Department of Energy in the United
States, the Commissariat \`a l'Energie Atomique and the Centre
National de la Recherche Scientifique / Institut National de Physique
Nucl\'eaire et de Physique des Particules in France, the Agenzia
Spaziale Italiana and the Istituto Nazionale di Fisica Nucleare in
Italy, the Ministry of Education, Culture, Sports, Science and
Technology (MEXT), High Energy Accelerator Research Organization (KEK)
and Japan Aerospace Exploration Agency (JAXA) in Japan, and the
K.~A.~Wallenberg Foundation, the Swedish Research Council and the
Swedish National Space Board in Sweden.\\ Additional support for
science analysis during the operations phase is gratefully
acknowledged from the Istituto Nazionale di Astrofisica in Italy and
the Centre National d'\'Etudes Spatiales in France.\\

\indent This publication makes use of molecular line data from the Boston
University-FCRAO Galactic Ring Survey (GRS). The GRS is a joint
project of Boston University and Five College Radio Astronomy
Observatory, funded by the National Science Foundation under grants
AST-9800334, AST-0098562, AST-0100793, AST-0228993 \& AST-0507657.
\end{acknowledgements}

\bibliographystyle{aa}
\bibliography{w49_hess_fermi}

\end{document}